\documentstyle[12pt,epsf]{article}

\begin{document}

\title{Derivation of ARCH(1) process from market price changes based on 
deterministic microscopic multi-agent}
\author{Aki-Hiro Sato$^{1}$\footnote{Electronic
mail:aki@sawada.riec.tohoku.ac.jp} and Hideki Takayasu$^{2}$ \\
Department of Applied Mathematics and Physics, \\
Kyoto University, Kyoto 606-8501, Japan, and \\
$^{2}$ Sony Computer Science Lab., Takanawa Muse Bldg., 3-14-13, \\
Higashi-Gotanda, Shinagawa-ku, Tokyo 141-0022, Japan.
}

\maketitle              

\begin{center}
{\bf Abstract}
\end{center}
{
A model of fluctuations in the market price including many deterministic 
dealers, who predict their buying and selling prices from the latest price 
change, is developed. We show that price changes of the model is 
approximated by 
ARCH(1) process. We conclude that predictions of dealers affected by 
the past price changes cause the fat tails of probability density function. 
We believe that this study bridges stochastic processes in econometrics with 
multi-agent simulation approaches.
}


\section{Introduction}
Market prices have been analyzed by many researchers. It is a famous result
that the probability density function of difference of the market prices 
follows the power law distribution (Mantegna et al. 1995). On the
other hand the stochastic processes of ARCH type are one of the most
exciting studies in econometrics. Statistical properties of ARCH
processes have been clarified by many 
researches. Namely a probability density of a dynamical variable of the 
ARCH process follows the power law distribution and its characteristics 
is applied to explain fluctuations in the real market (Engle 1982, 
Bollerslev 1986, Nelson 1990). Recently economists and physicists are
interested in multi-agent simulations of problems motivated economically
(Takayasu et al. 1992, Bak et al. 1997, Sato et al. 1998, Johnson et
al. 1998, Lux et al. 1999). This approach is to study an artificial
market in which programming agents sell and buy their stocks and
actually market price changes can be calculated by an interaction of
many dealers.

\begin{figure}[h]
\begin{center}
\epsfxsize=220pt
\epsfbox{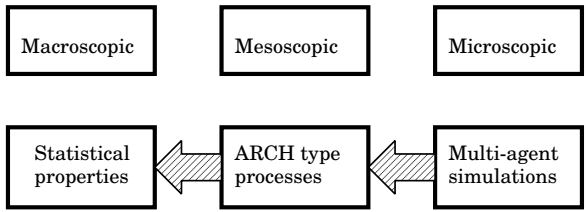}
\end{center}
\caption{The conceptual illustration of correspondence of scope to
study methods.}
\label{fig:macro-micro}
\end{figure}

On the viewpoint of statistical mechanics we can consider investigating 
statistical properties of market prices, the stochastic processes of the 
ARCH type and multi-agent simulations to be macroscopic, mesoscopic and 
microscopic, respectively. (see fig. \ref{fig:macro-micro}). 
Our study will bridge the multi-agent simulation with stochastic processes
of the ARCH type.

In this article we develop a simple market model with many 
deterministic dealers. The dealers estimate their buying and selling
prices from the latest change of the market price. We show 
that from market price changes of the
model a stochastic process of the ARCH type can be derived. Let us give 
a brief outline of this article. In the second chapter we describe the
process of ARCH type. In the third chapter we mention a simple model
of fluctuations in market price based on dealers. In the fourth chapter 
we analytically derive an ARCH(1) process for price changes from the 
proposed model. The fifth chapter is devoted to the 
concluding remarks.

\section{The ARCH (q) processes}
The ARCH abbreviates autoregressive conditional heteroskedasticity, 
which has been introduced by Engle in econometrics (Engle 1982).
The ARCH(q) process is formalized by 
\begin{equation}
\left\{
\begin{array}{lcl}
\epsilon_{s} &=& \sigma_{s}\cdot Z_{s} \\
\sigma_{s}^2 &=& \alpha_0 + \alpha_1\epsilon_{s-1}^2 + \cdots + 
\alpha_{s-q}\epsilon_{s-q}^2
\end{array}
\right.,
\label{eq:ARCH-q}
\end{equation}
where $\epsilon_s$ is an interesting variable, $Z_s$ is a stochastic variable,
$\sigma_s$ is called volatility. $\alpha_i$ ($i = 0 \ldots q$) is a positive 
parameter. Consider $q=1$ as the most simple case of the ARCH(q) stochastic
process. It seems sufficient to discuss the case of $q=1$ in order to
see statistical properties of the ARCH process. Especially when $q=1$ 
eq. (\ref{eq:ARCH-q}) is written by
\begin{equation}
\left\{
\begin{array}{lcl}
\epsilon_{s} &=& \sigma_{s}\cdot Z_{s} \\
\sigma_{s}^2 &=& \alpha_0 + \alpha_1\epsilon_{s-1}^2.
\end{array}
\right..
\label{eq:ARCH}
\end{equation}
Eliminating $\sigma_s$ into eq. (\ref{eq:ARCH}) one can get
an alternative expression, 
\begin{equation}
\epsilon_{s} = \sqrt{\alpha_0 + \alpha_1 \epsilon_{s-1}^2} \cdot Z_s.
\label{eq:alternate-ARCH}
\end{equation}
Moreover eq. (\ref{eq:alternate-ARCH}) is approximated as
\begin{equation}
\epsilon_s =
\left\{
\begin{array}{ll}
\sqrt{\alpha_0} \cdot Z_s & (|\epsilon_{s-1}| \leq \sqrt{\alpha_0/\alpha_1}) \\
\sqrt{\alpha_1} \cdot |\epsilon_{s-1}| \cdot Z_s & 
(|\epsilon_{s-1}| > \sqrt{\alpha_0/\alpha_1}) 
\end{array}
\right..
\end{equation}
This equation is equivalent to a random multiplicative process 
(Takayasu et al. 1997). Using the random multiplicative process theory
(Sato et al. 2000) it is easy to prove that a probability density function
of $\epsilon_s$ has power law tails,
\begin{equation}
p(\epsilon) \propto |\epsilon|^{-\beta-1},
\end{equation}
where $\beta$ is given by
\begin{equation}
\alpha_1^{\beta/2} \langle |Z|^{\beta} \rangle = 1.
\end{equation}

\section{Dealer model} 
\begin{figure}[h]
\begin{center}
\epsfxsize=220pt
\epsfbox{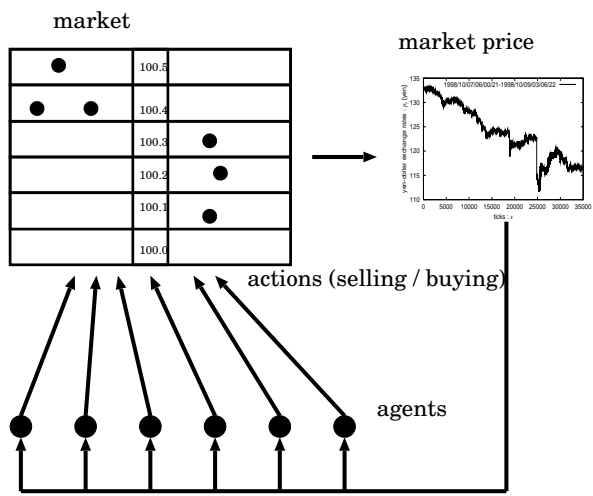}
\end{center}
\caption{The conceptual illustration of dealer model. The inputs of the 
market are orders from dealers. The output of the market is a market price. 
The input of an agent is a sequence of latest price changes. The output of 
an agent is a selling or a buying.}
\label{fig:concept}
\end{figure}

We show a brief explanation of the market model consisted of many simple 
deterministic dealers. As shown in fig. \ref{fig:concept} this model can 
be separated into two parts; One is a market mechanism, which describes 
how to determine a market price from orders. The other is an algorithm 
of agents, which governs how to order a selling or buying on the market and 
how to modify their bid prices. In the following subsections we
explain these parts, initial conditions and parameters and also
results of numerical simulations.

\subsection{Market mechanism}
Suppose that a market of a competitive buying and selling in which 
$N$ dealers give their orders for a common order board. $S_i(t)$ and 
$B_i(t)$ represent a selling price and buying price for the $i$th dealer 
at time $t$, respectively. It is assumed that buying prices
and selling prices individually compete in this market. Namely the
maximum buying price and the minimum selling price are effective in
the market. Thus the condition for a trade to occur is given by the 
inequality,
\begin{equation}
L(t) = \max\{B(t)\} - \min\{S(t)\} \geq 0,
\end{equation}
where $\max\{B(t)\}$ represents the maximum buying price in all the
buyers, and $\min\{S(t)\}$ the minimum selling price in all the sellers. 
Suppose that the market price $P(t)$ is determined as an arithmetic 
mean of the selling price and the buying one when the transaction 
occurs. Otherwise the latest market price is maintained. Namely, 
\begin{equation}
P(t) = \left\{
\begin{array}{ll}
\frac{1}{2}\bigl(\max\{B\} + \min\{S\}\bigr) &
\mbox{($L(t) \geq \Lambda$)}
\\ 
P(t-1) &
\mbox{($L(t) < \Lambda$)} 
\end{array}
\right..
\label{eq:market-mechanism}
\end{equation}

\subsection{Dealer algorithm}
We suppose that a seller goes on decreasing his expectation of selling
price until he can sell a stock and that a buyer goes on increasing his 
expectation of buying price until he can buy a stock. A rule to modify 
his expecting price is given by
\begin{equation}
B_i(t+1) = B_i(t) + |1+c_i\Delta P_{prev}|a_i(t),
\label{eq:dealers-rule}
\end{equation}
where $\Delta P_{prev}$ denotes the latest change of the market price. 
From the above assumption for the seller and the buyer the $i$th dealer
is a seller when $a_i(t)<0$ and he is a buyer when $a_i(t)>0$. A term of 
$|1+c_i\Delta P_{prev}|$ means that modification of his expectation of 
price depends on the latest market price change. Here $c_i$ is a prediction
coefficient dependent on the dealer. 

Suppose that all the dealers have a small asset. Thus
it means that a seller/buyer changes his position into a buyer/seller after
a trade. Because of the assumptions that each dealer keeps his position 
till a trading, an evolution rule of $a(t)$ is given by
\begin{equation}
a_i(t+1) = \left\{
\begin{array}{ll}
-a_i(t) & \mbox{\small (a seller and a buyer when a trade occurs)}
\\
a_i(t)  & \mbox{\small (otherwise)}
\end{array}
\right..
\end{equation}

\subsection{Initial conditions and parameters}
$B_{i}(0)$, $a_{i}(0)$ and $c_i$ are given by random numbers of which 
a range is $[-\Lambda/2,\Lambda/2]$, $[-\alpha,\alpha]$ and $[-c*,c*]$, 
respectively. We put $P(0)=0$ and $\Delta P_{prev}=0$. The dealer's 
rule is deterministic except initial conditions. This model has four 
parameters; the dealer number $N$, $\alpha$ for $a_i(t)$, difference 
between selling and buying price $\Lambda$ and $c*$ for $c_i$.

\subsection{Numerical Simulation}
\begin{figure}[hbt]
\begin{center}
\epsfxsize=220pt
\epsfbox{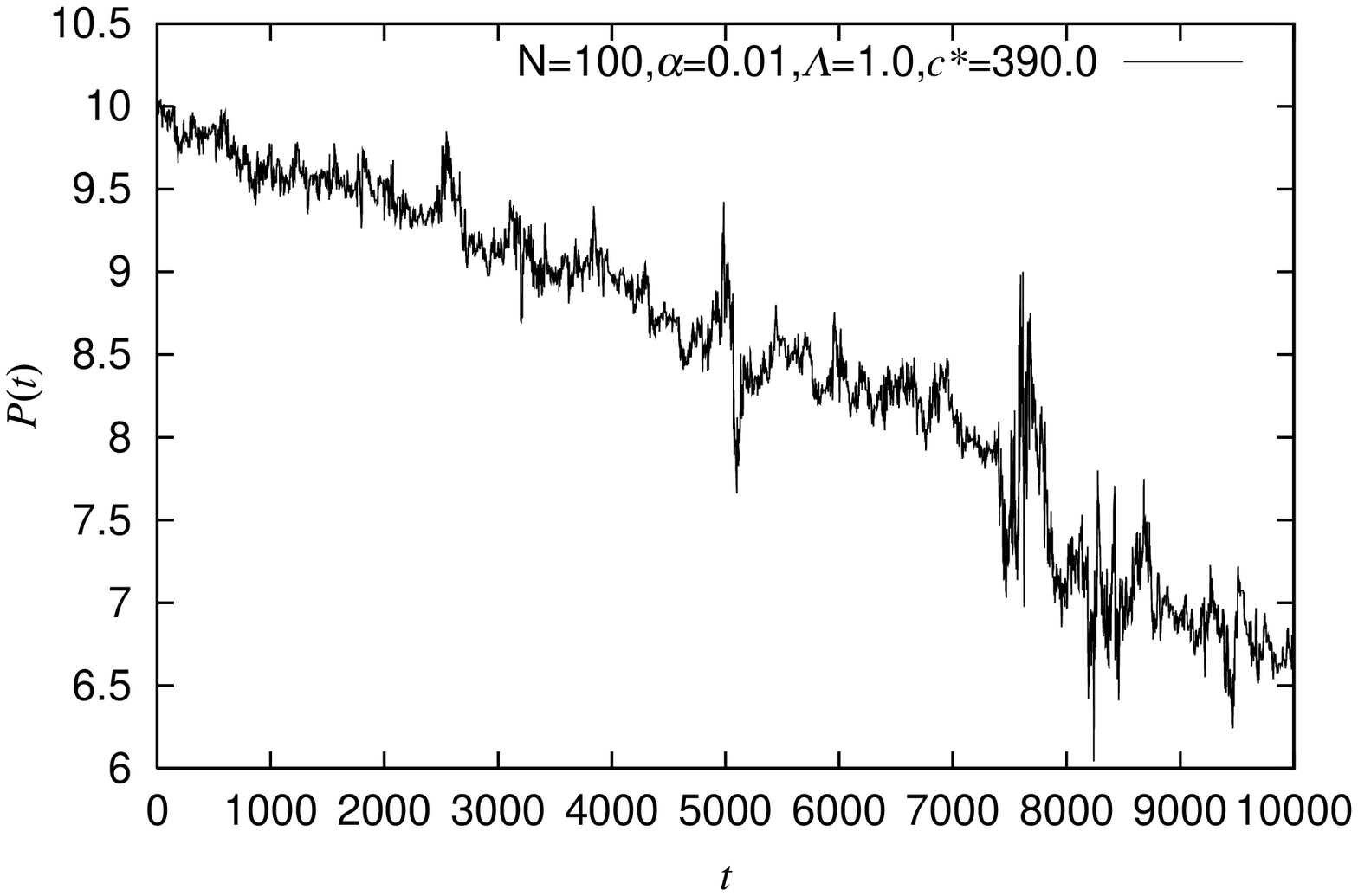}
\epsfxsize=220pt
\epsfbox{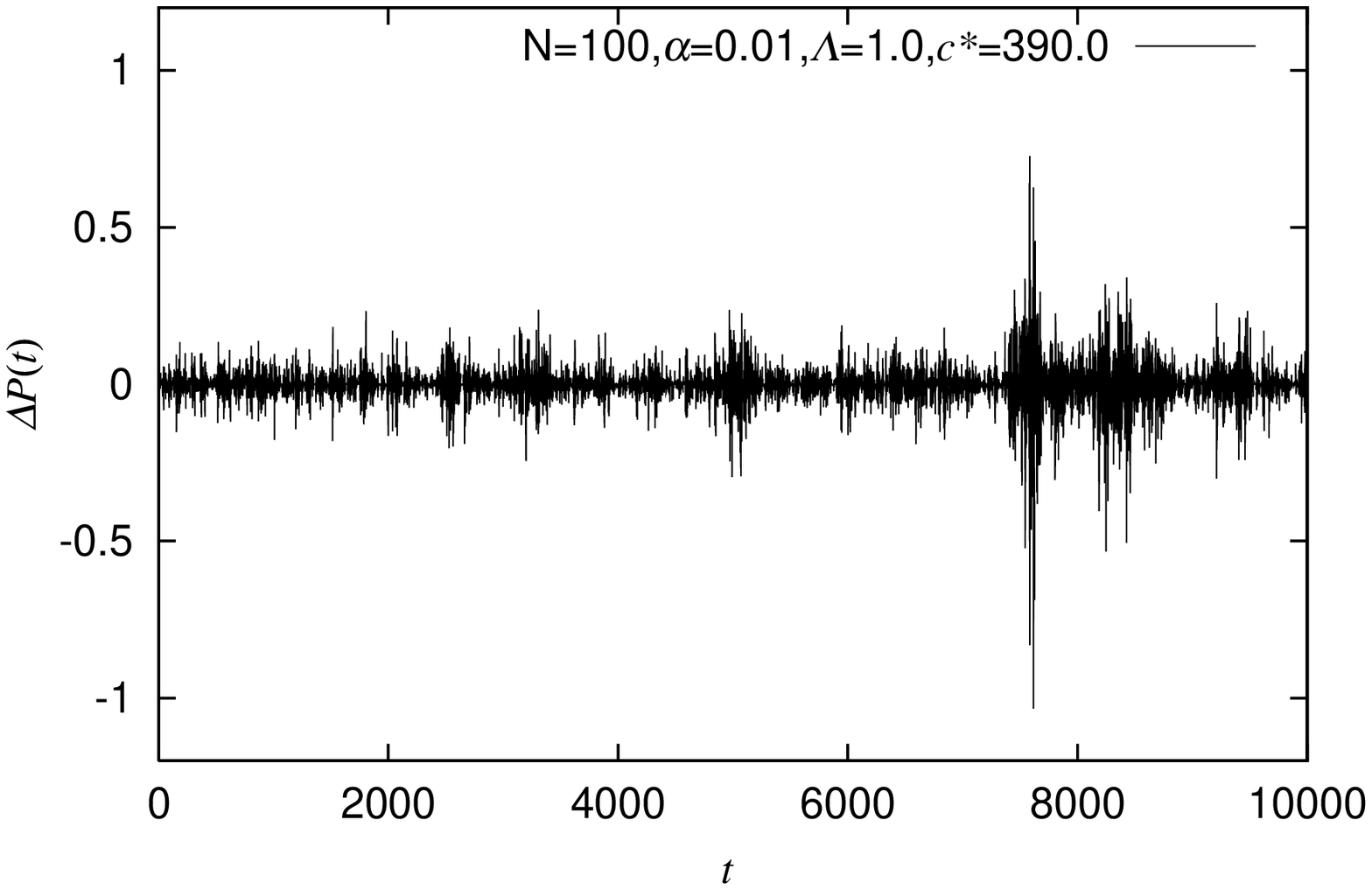}
\end{center}
\caption[]{Time series of artificial prices (right). Time series of changes 
of the artificial prices (left). X axis represents a step $t$. Parameters are
$N=100$,$\alpha=0.01$,$\Lambda=1.0$ and $c^{*}=390.0$.}
\label{fig:price-seq}
\end{figure}

\begin{figure}[hbt]
\begin{center}
\epsfxsize=220pt
\epsfbox{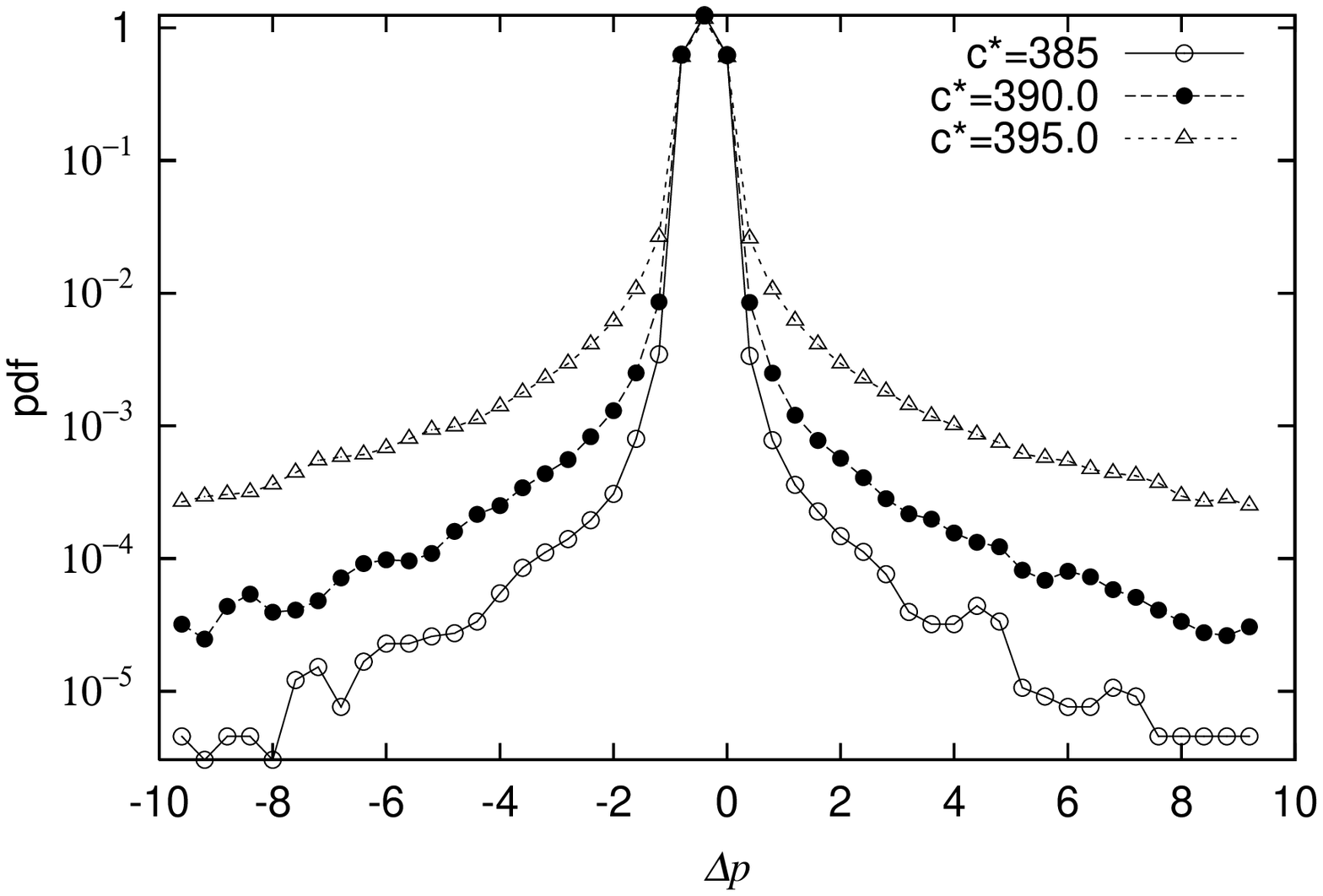}
\epsfxsize=220pt
\epsfbox{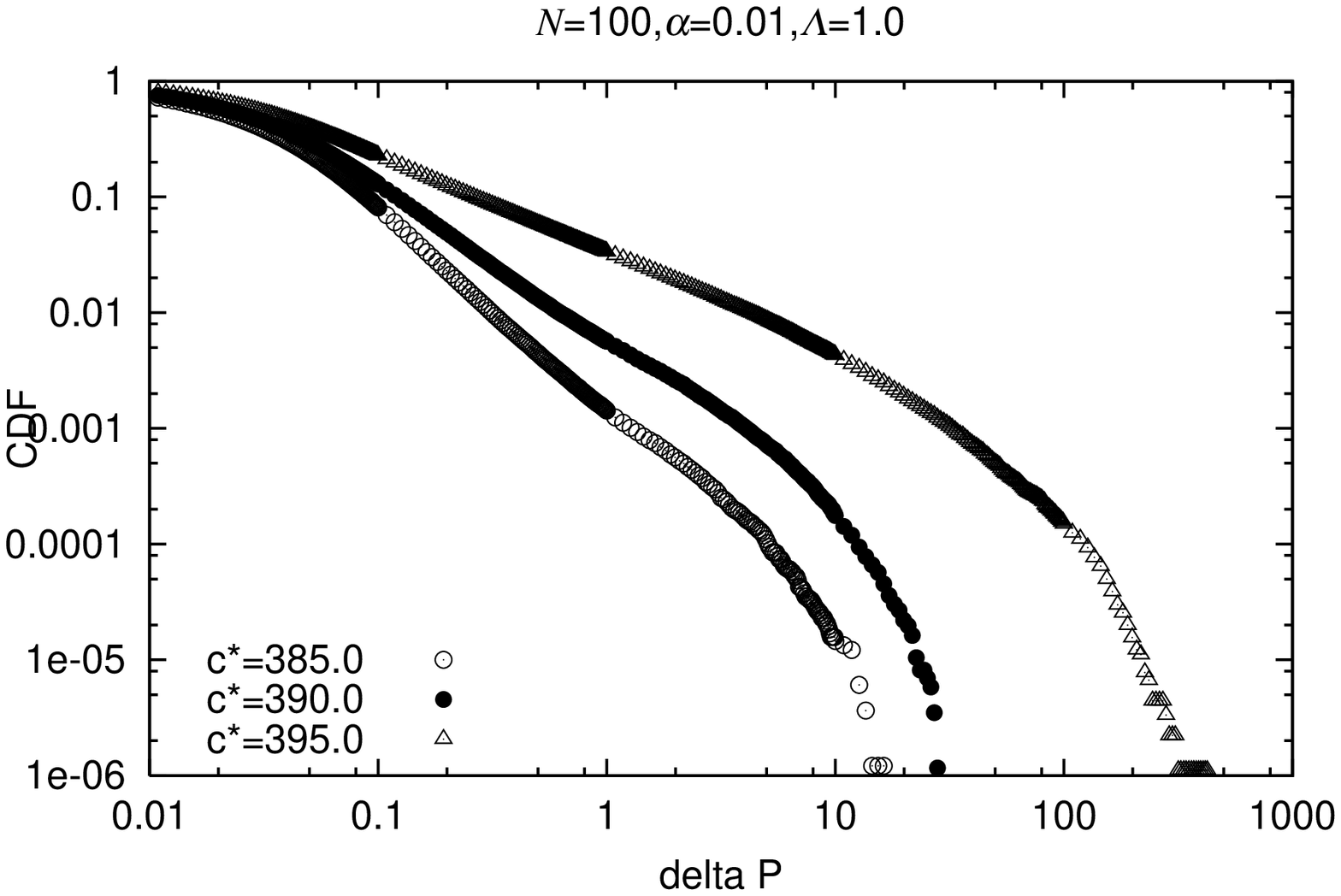}
\end{center}
\caption{Semi-log plots of the probability density functions of the 
changes (right). Log-log plots of the cumulative distributions of the 
changes (left). Parameters are $N=100$,$\alpha=0.01$,$\Lambda=1.0$.}
\label{fig:pdfs-cdfs}
\end{figure}

Fig. \ref{fig:price-seq}
show a typical example of time series of market price
$P(t)$ simulated numerically with price changes $\Delta P(t) 
= P(t) - P(t-1)$. Because an event of a change of the market price 
occurs in Poisonian 
we define a jump $\Delta p$ as a market price change on which a trade occurs.
Fig. \ref{fig:pdfs-cdfs} shows probability density functions (PDF) of
the jumps and their corresponding cumulative distribution functions
(CDF), which is defined by 
\begin{equation}
P(\geq|x|) = \int_{-\infty}^{-|x|}p(x')dx' + \int_{|x|}^{\infty}p(x')dx'.
\end{equation}
From fig. \ref{fig:pdfs-cdfs} we can find linear part meaning that 
CDFs follow power law distributions of the following form 
\begin{equation}
P(\geq|x|) \propto |x|^{-\beta},
\end{equation}
where $\beta$ is a power law exponent. Moreover the power law exponent
obviously depends on a value of $c^{*}$. 

\section{From Microscopic to mesoscopic}
\begin{figure}[hbt]
\begin{center}
\epsfxsize=220pt
\epsfbox{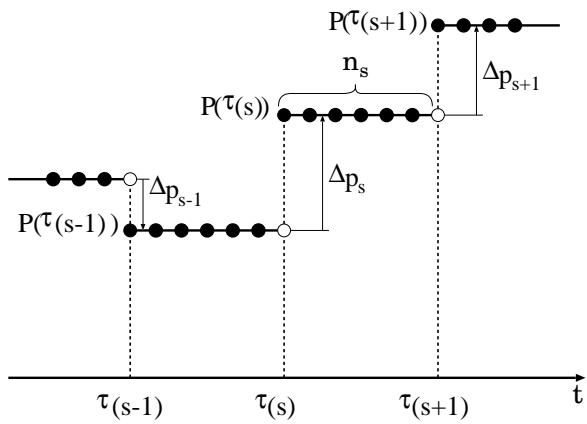}
\end{center}
\caption{A conceptual illustration of a temporal development of the market
price $P(t)$. $\tau(s)$ represents a step for the $s$th trade to occur.
$\Delta p_s$ is a price change on that step. $n_s$ denotes steps between
the $s-1$th trade and the $s$th.}
\label{fig:fluctuation}
\end{figure}

As shown in fig. \ref{fig:fluctuation} $\Delta p_s$ represents a
change of the market price on the $s$th trade, and $n_s$ a time interval
between the $s$th trade and the $s+1$st. We define $M_s$ as
the buying price at the $s$th trade and $m_s$ as the selling price,
respectively. From the assumptions of eq. (\ref{eq:market-mechanism})
$\Delta p_s$ can be estimated as 
\begin{eqnarray}
\nonumber
\Delta p_s &=& \frac{1}{2}(m_s + M_s + \Lambda) -
\frac{1}{2}(m_{s-1}+M_{s-1} + \Lambda) 
\\
&=& \frac{1}{2}(m_s - m_{s-1}) + \frac{1}{2}(M_s - M_{s-1}).
\label{eq:Delta-p}
\end{eqnarray}
The first term and the second term on the second line are approximated by
\begin{eqnarray}
m_s - m_{s-1} &=& |1+c_i\Delta p_{s-1}| a_{i} n_{s-1}, 
\label{eq:delta-m}
\\ 
M_s - M_{s-1} &=& |1+c_j\Delta p_{s-1}| a_{j} n_{s-1},
\label{eq:delta-M}
\end{eqnarray}
where the subscript $i$ denotes the dealer who gives the lowest selling
price, and $j$ the dealer who gives the highest buying one. 
Substituting eqs. (\ref{eq:delta-m}) and (\ref{eq:delta-M}) into 
eq. (\ref{eq:Delta-p}) yields 
\begin{equation}
\Delta p_s = \frac{1}{2}|1+c_i\Delta p_{s-1}|a_i n_{s-1} + \frac{1}{2}
|1+c_j\Delta p_{s-1}|a_j n_{s-1}.
\label{eq:Delta-p-rewrite}
\end{equation}

When the dealer number $N$ is large it can be assumed that dealers'
prices of expectation are distributed uniformly. By taking square of 
eq. (\ref{eq:Delta-p-rewrite}) for all the dealers and averaging over 
ensemble under the condition that $\Delta p_{s-1}$ is realized, we get 
the following equation,
\begin{equation}
\langle \Delta p_s ^2\rangle = \frac{1}{2}\langle a^2
\rangle (1 + \langle c^2 \rangle \Delta p_{s-1}^2)\langle n_{s-1}^2 \rangle.
\label{eq:Delta-p2}
\end{equation}
From the assumptions of $c_i$ and $a_i$ we have $\langle c \rangle = 0$, 
$\langle c^2 \rangle = c*^2/3$, $\langle a \rangle = 0$ and 
$\langle a^2 \rangle = \alpha^2/3$. Assuming that 
$\langle \Delta p \rangle = 0$, eq. (\ref{eq:Delta-p2}) becomes, 
\begin{equation}
\langle \Delta p_s ^2 \rangle = \alpha^2
(1 + \frac{c*^2}{3} \Delta p_{s-1}^2)\langle n_{s-1}^2 \rangle.
\label{eq:ARCH-for-deltap}
\end{equation}

Eq. (\ref{eq:ARCH-for-deltap}) shows that the variance of $\Delta p_s$ relates
to a realized value on $s-1$ and it has the same form with eq. 
(\ref{eq:ARCH}) without the term of $\langle n_{s-1}^2 \rangle$. 
Supposing that $\sigma_{s}'$ is a stochastic variable with a zero-average 
and a normal variance we can also rewrite eq. (\ref{eq:ARCH-for-deltap}) as
\begin{equation}
\Delta p_s = \sqrt{1+\frac{c*^2}{3} \Delta
p_{s-1}^2}\alpha\sqrt{\langle n_{s-1}^2 \rangle}\sigma_{s}'.
\label{eq:dealer-ARCH}
\end{equation}
Eq. (\ref{eq:dealer-ARCH}) becomes identical to eq.
(\ref{eq:alternate-ARCH}) with the relations $\alpha_0=1$, $\alpha_1=c*^2/3$ 
and $\sigma_s = \alpha \sqrt{\langle n_{s-1}^2 \rangle} \sigma_{s}'$.

\section{Conclusions}
We introduced the model of an artificial market with many deterministic 
dealers and showed the outline of the stochastic process of ARCH 
type. The ARCH(1) process can be derived from the changes of 
the market price in this model. We conclude that the fluctuation
of the market price is approximated by the ARCH type process if 
each dealer changes his expectation of a buying/selling price 
proportional to the lastest market price change. We expect 
that our approach will bridge the stochastic processes of the 
ARCH type in econometrics with dynamical market models consisted of
dealers, and our understanding about the basic properties of markets 
will be deepened.


\end{document}